\documentclass[a4paper]{jpconf}
\usepackage{graphicx}
\usepackage{amsmath}
\usepackage[]{hyperref}
\hypersetup{
	pdftitle    = Cavity-Based Charge Diagnostic
	pdfauthor   = Simon Bohlen,
}
\usepackage[separate-uncertainty]{siunitx}
\usepackage{fancyhdr}

\begin{document}

\title{Noninvasive cavity-based charge diagnostic for plasma accelerators}

\author{S.~Bohlen$^{1,2}$, O.~Kononenko$^{1}$, J.-P.~Schwinkendorf$^{1,2}$, F.~Gr\"uner$^{1,2}$, D.~Lipka$^{1}$, M.~Meisel$^{1,2}$, C.A.J.~Palmer$^{1}$, T.~Staufer$^{2}$, K.~P\~oder$^{1}$ and~J.~Osterhoff$^{1,2}$}

\address{$^{1}$Deutsches Elektronen-Synchrotron DESY, Notkestr. 85, 22607 Hamburg, Germany\\
$^{2}$Universit\"at Hamburg, Luruper Chaussee 149, 22607 Hamburg, Germany}

\ead{simon.bohlen@desy.de}

\pagestyle{fancy}
\fancyhead{}
\fancyfoot{}
\renewcommand{\headrulewidth}{0pt}
\fancyfoot[R]{\thepage}

\begin{abstract}
The charge contained in an electron bunch is one of the most important parameters in accelerator physics. Several techniques to measure the electron bunch charge exist. However, many conventional charge diagnostics face serious drawbacks when applied to plasma accelerators. For example, integrating current transformers (ICTs or toroids) have been shown to be sensitive to the electromagnetic pulses (EMP) originating from the plasma, whereas scintillating screens are sensitive to background radiation such as betatron radiation or bremsstrahlung and only allow for a destructive measurement of the bunch charge. We show measurements with a noninvasive, cavity-based charge diagnostic (the DaMon), which demonstrate its high sensitivity, high dynamic range and resistance towards EMP. The measurements are compared to both an ICT and an absolutely calibrated scintillating screen. 
\end{abstract}

\section{Introduction}
Plasma acceleration driven by laser pulses (LPA) \cite{TajimaDawson} or particle beams (PWFA) \cite{Chen1985} enables the acceleration of electron bunches with gradients several orders of magnitude higher than in conventional radio-frequency  (RF) accelerators. Thus, electron energies of hundreds of MeV can be achieved after a few millimetres of acceleration. While a lot of progress has been made in terms of stabilisation and control of plasma accelerators \cite{Khz,BohlenPRAB,Maier2020,emittancepreservation}, the level of stability can not yet compete with their RF counterparts. As such, the accurate single-shot detection and characterisation of electron bunches from plasma accelerators is often desirable. Furthermore, accurate and non-invasive diagnostics are required when plasma accelerators are used in applications such as advanced light sources based on Thomson scattering \cite{Brümmer2022} or when driving free electron lasers \cite{Wang2021}. Integrating current transformers (ICTs), which are commonly used in conventional accelerators, `\textit{have become the instrument of choice for measuring total charge from plasma-based accelerators}' \cite{Downer2018}, even though they have early been identified to drastically overestimate the charge in plasma accelerators by up to an order of magnitude \cite{Downer2018,Glinec2006,Hidding2007}. The overestimation of the measured charge using ICTs originates from electromagnetic pulses (EMP) present in plasma accelerators. As such, accurate determination of the charge in proximity of strong EMP has often only been possible using destructive methods, such as absolutely calibrated scintillating screens or Faraday cups. 

Here, we demonstrate the use of a non-invasive, absolutely calibrated, cavity-based charge diagnostic called DaMon, which offers a precise measurement of the electron bunch charge over a wide range. In addition, the DaMon is shown to be resilient to EMP, even in the presence of a high-voltage discharge in an active plasma lens (APL). It is further compared to an ICT and an absolutely calibrated scintillating screen (type DRZ-high). In these comparisons, we find an overestimation of the charge by the ICT, especially at low electron bunch charges, which is in line with previous findings. In contrast, the DaMon is shown to be able to measure the bunch charge over a wide dynamic range of several orders of magnitude in a non-invasive way. Due to its low noise level, it was found to be more sensitive to low-charge electron bunches than the other two diagnostics in the comparison and enabled the detection of bunches with charges as low as tens of femtocoulomb.

\section{Cavity-based charge detector: The DaMon}
The DaMon is a cavity that non-invasively measures the charge of the electron bunch. Originally, it was developed to measure bunch charges and dark currents at the accelerators FLASH and XFEL at DESY \cite{Lipka2011, Lipka2013}. 
The DaMon is a resonator made from stainless steel with the frequency of the first monopole mode (TM01) at 1.3 GHz. An electron bunch passing through the cavity will induce the TM01 mode with a voltage of:

\begin{equation}
	\label{Eq:Voltage_TM01}
	U = U_0\;\sin(\omega t)\;e^{-\frac{t}{\tau}},
\end{equation}
where $\omega$ = 2$\pi f$ with $f$ being the resonance frequency of \SI{1.3}{GHz}, and $\tau$ being the decay time given by $\tau = Q_L/(\pi f)$ where $Q_L$ is the resonator loaded quality factor. 
The amplitude of the induced voltage is given by:
\begin{equation}
	\label{Eq:U0_prop_q}
	   U_0 = q\;S.
\end{equation}
Here, q is the electron bunch charge and S is the sensitivity defined as $S = \pi f \sqrt{\frac{Z}{Q_{ext}}(\frac{R}{Q})}$, where $Z$, $Q_{ext}$ and $R/Q$ are the line impedance, external quality factor and simulated normalized shunt impedance, respectively. The values for $Z$ and $Q_{ext}$ can be determined by independent calibration measurements such that the electron bunch charge can be calculated using Eq. \ref{Eq:U0_prop_q} by measuring the amplitude of the induced TM01 mode without further calibration. Two antennae are used for coupling-out the signal from the cavity. The signal from each of the two antennae is processed by electronics which consist of circulator, band-pass filter, logarithmic detector and offset and gain control. One channel is equipped with limiter and down conversion to an intermediate frequency, which enables higher sensitivity for this branch. Due to the logarithmic output and two channels with different sensitivity, the DaMon enables charge detection with a dynamic range of seven orders of magnitude. More detailed information on the DaMon can be found in the original publications \cite{Lipka2011, Lipka2013} and theses which describe its first use with plasma accelerators \cite{LenaPHDThesis,JPThesis,SimonThesis}.


\section{Experimental setup}
\begin{figure*}[h!tb]
	\centering        
	\includegraphics[width=0.85\textwidth]{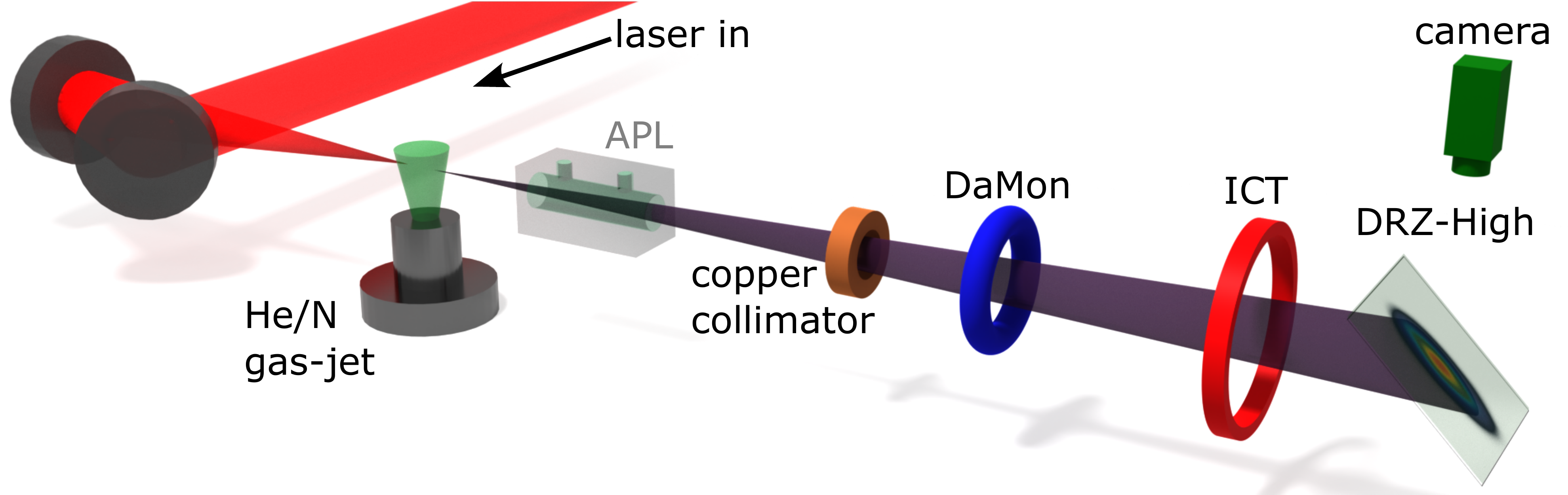}
    \caption[ExpSetup]{Schematic of the experimental setup.}
	\label{fig:ExpSetup}
\end{figure*}

The experiments for the tests and comparison of charge diagnostics were conducted at the FLARE facility at DESY \cite{BohlenPRAB,BohlenPRL}. A schematic of the experimental setup is depicted in Fig. \ref{fig:ExpSetup}. The laser was focused onto a \SI{1}{\mm} supersonic gas jet using an f/12 off-axis parabola. The gas used was 99.5 \% helium with 0.5 \% nitrogen doping to accelerate electrons via ionisation injection (II) \cite{Chen2006,Oz2007,Pak2010,McGuffey2010,Clayton2010}.

To measure the electron bunch charge the DaMon, an ICT, and a scintillating screen (type DRZ-High) were used. The DRZ screen was absolutely calibrated at the Elbe accelerator at Helmholtz-Zentrum Dresden-Rossendorf prior to the experiments \cite{Kurz2018, Schwinkendorf2019} and cross-calibrated using tritium sources \cite{SimonThesis}. While both tritium sources and the used DRZ screen emit most of their light in a similar wavelength regime, differences in the emitted spectrum prevented the use of narrow-band interference, notch or edge-band filters for imaging of the DRZ screen as this would have affected the absolute calibration. A background subtraction was applied for measurements using the DRZ screen, which consisted of 100 background images without any laser light.

The three charge diagnostics were installed at distances of approximately \SI{1.05}{m}, \SI{1.3}{m} and \SI{1.4}{m} from the plasma source for DaMon, ICT and DRZ screen, respectively. To ensure that all three diagnostics measured the same amount of electrons, a copper collimator was installed between the electron source and the charge diagnostics. The collimator reduced the available beam aperture to \SI{30}{mm} at a distance of approximately \SI{0.9}{m}. The diagnostic with the smallest aperture of \SI{40.5}{mm} was the DaMon, whereas ICT and DRZ screen had an aperture of \SI{72}{mm} and \SI{65}{mm}, respectively. As such, the maximum beamsize was limited by the aperture of the copper collimator and not the apertures of any of the diagnostics. For parts of the experiments an active plasma lens (APL) \cite{Tilborg2015} was placed \SI{0.15}{m} behind the source to study the influence of strong EMP on the diagnostics.

A measurement of the electron spectra during the charge calibration was not possible, due to the detection of the electron bunches on the DRZ screen. Previously, broadband electron bunches with peaks around \SIrange{50}{80}{MeV} were measured at this setup. At these energies, the response of the detectors should be energy independent. However, electrons with energies of less than a few MeV could deposit more energy in the scintillating screen compared to the more energetic electrons \cite{Glinec2006}. 
More information on the general setup and further details on electron parameters previously measured can be found in other publications \cite{BohlenPRAB,BohlenPRL}.

\section{Comparison of DaMon, ICT and DRZ screen}
In this section, we provide a comparison of the three independently absolutely calibrated diagnostics DaMon, ICT, and DRZ screen. In the first part, the sensitivity to noise from plasma creation and its effect on measuring low charges is investigated. In the second part, we further explore the impact of intense EMP on the charge measurement of the non-invasive diagnostics DaMon and ICT. In the end, an overview of the performance of all three diagnostics is presented to investigate their capability to detect charges spanning multiple orders of magnitude.

\subsection{Sensitivity to low charges}
\begin{figure*}[h!tb]
	\centering        
	\includegraphics[width=15cm]{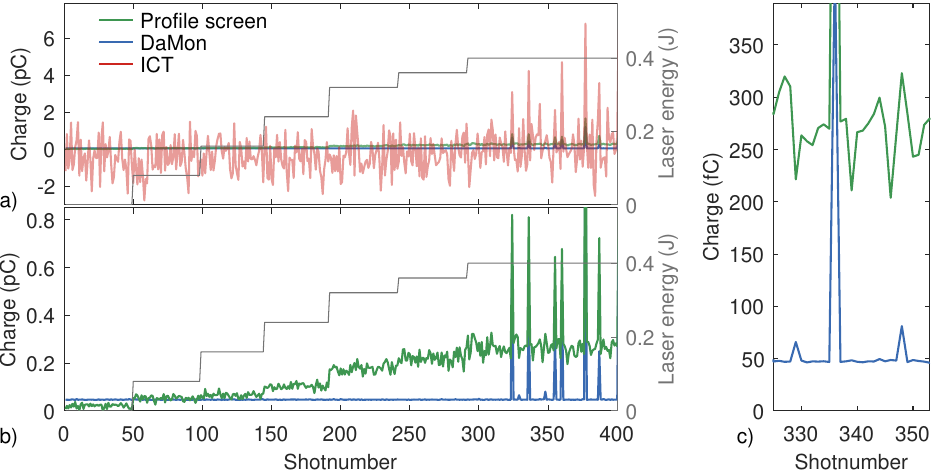}
	\caption[LowCharge]{a) Comparison of the three charge diagnostics at low laser energies. b) Same as panel a), but without the ICT. c) Zoom into shots 325-353 of panel b). The grey lines in panels a) and b) indicates the laser energy.}
	\label{fig:SensitivityLowCharge}
\end{figure*}

To investigate the behaviour of the diagnostics in an environment with EMP and at low charges, studies were performed below and around the ionisation injection threshold. The results are depicted in Fig.~\ref{fig:SensitivityLowCharge}.

In these scans, the laser energy was increased in steps with increasing shotnumber, as indicated by a grey line in the panels Fig.~\ref{fig:SensitivityLowCharge}a and Fig.~\ref{fig:SensitivityLowCharge}b. The threshold for ionisation injection in the used setup was at a laser energy of around \SI{400}{mJ}. Therefore, the first 300 laser shots enable plasma generation and its related background effects to be studied on the charge diagnostics without acceleration of charged particles. From Fig.~\ref{fig:SensitivityLowCharge}a, it is visible that the ICT has a noise level of approximately \SI{\pm 2}{pC} and therefore is not suited for the reliable detection of low charges. The DaMon has a constant noise level of \SI{47 \pm 1}{fC} independent of laser energy, demonstrating its resistance to EMP noise from the plasma generation as is visible in Fig.~\ref{fig:SensitivityLowCharge}b. The camera looking at the DRZ screen shows an increase in signal with increasing laser energy, which could be interpreted as charges of up to \SI{250}{fC} at the injection threshold of \SI{400}{mJ} laser energy. However, this signal likely originates from laser light and/or plasma light as the more sensitive DaMon does not see this step wise ramp up in charge and no accelerated electrons would be expected at the used laser energies. This potential background signal can not be accounted for in the background images without laser light as it is dependent on laser energy. While the imaging system of the DRZ screen has been carefully shielded, the additional use of notch filters was not possible due to the cross calibration using tritium sources as described above.

When the laser energy reached the threshold of \SI{400}{mJ}, the generation of electron bunches with low charge was possible. The ability of the DaMon to clearly distinguish these bunches from the very constant background is visible in Fig.~\ref{fig:SensitivityLowCharge}c, where bunches with charges of \SI{66}{fC} and \SI{81}{fC} are clearly distinguishable form the background level of \SI{47}{fC}. At the same time, no signal above the noise level gets detected by the DRZ screen for these two shots. The same is true for the ICT, which is not able to resolve charges this low. The ICT even fails to detect some bunches, where both DaMon and DRZ screen have signals of about \SI{300}{fC} above noise level, such as shot 355, as can be seen in Fig.~\ref{fig:SensitivityLowCharge}a and Fig.~\ref{fig:SensitivityLowCharge}b.

The noise level of the DaMon is currently limited by the electronics and could be lowered to a few femtocoulomb or even attocoulomb levels with modified and cooled electronics \cite{Lipkaprivatecommunication}. In the past, charges as low as \SI{25}{fC} have been measured using the DaMon in a different configuration at the same setup \cite{LenaPHDThesis}. The sensitivity of the scintillating screens could also be increased by optimising camera and imaging system, while careful shielding from plasma and laser light would be required as demonstrated here. The ICT used in this study is not suitable for reliable detection of low charge electron bunches. For low charges, Turbo-ICTs could be used which have a lower noise level of approximately \SI{1}{pC} \cite{Nakamura2016}.

\subsection{Sensitivity to EMP}
\begin{figure*}[h!tb]
	\centering        
	\includegraphics[width=12cm]{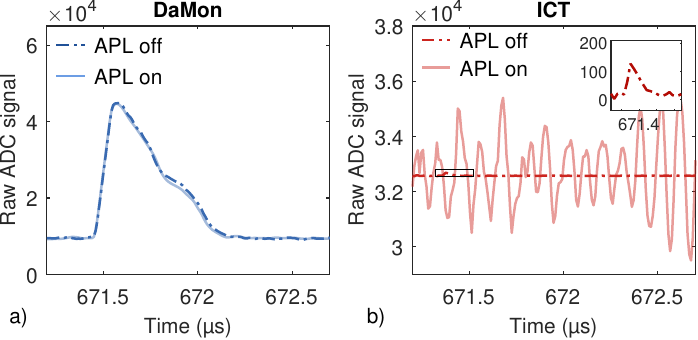}
	\caption[DaMonProp]{Comparison of the raw ADC traces for a) DaMon with APL off (dark blue, dashed) and APL on (light blue, solid) and b) ICT with APL off (dark red, dashed) and APL on (light red, solid). The two traces with and without the APL discharging shown in each panel are simultaneous measurements of the same bunch. The inset in panel b) shows the ADC trace from the ICT without the APL. For the inset, the minimum trace value was subtracted to show the magnitude of the raw signal without the APL.}
	\label{fig:DaMonPropaganda}
\end{figure*}

The effect of plasma EMP on the two non-invasive diagnostics DaMon and ICT was studied in more detail using an active plasma lens (APL). The raw analog-to-digital converter (ADC) traces of an electron bunch charge measurement with and without the APL discharging are depicted in Fig.~\ref{fig:DaMonPropaganda} for both DaMon and ICT. 

In Fig.~\ref{fig:DaMonPropaganda}a, it is visible that the raw ADC trace of the DaMon does not seem to be affected by the EMP signal that is induced from discharging the APL. In contrast, the signal picked up by the ICT when using the APL is orders of magnitude higher than the raw ICT signal of an electron bunch passing through without the APL discharging as can be seen in Fig.~\ref{fig:DaMonPropaganda}b. 
While it is possible that parts of the EMP noise visible in the ADC trace of the ICT are induced in the cables rather the device itself, it is obvious that careful implementation and shielding of an ICT is required in environments where EMP noise is present. Without such shielding, ICTs can overestimate the charge when installed close to the plasma source, as has been shown by other groups before \cite{Glinec2006, Hidding2007}. In the past, shielding from EMP was for example achieved by placing thin foils inside the beamline and installing the ICT several meters away from the plasma source \cite{Nakamura2016,Nakamura2011}. The DaMon on the other hand is unaffected by EMP noise, making it a suitable diagnostic for non-invasive charge measurements of plasma accelerators even in close proximity to the plasma source.

\subsection{High dynamic range charge detection}
\begin{figure*}[h!tb]
	\centering        
	\includegraphics[width=16cm]{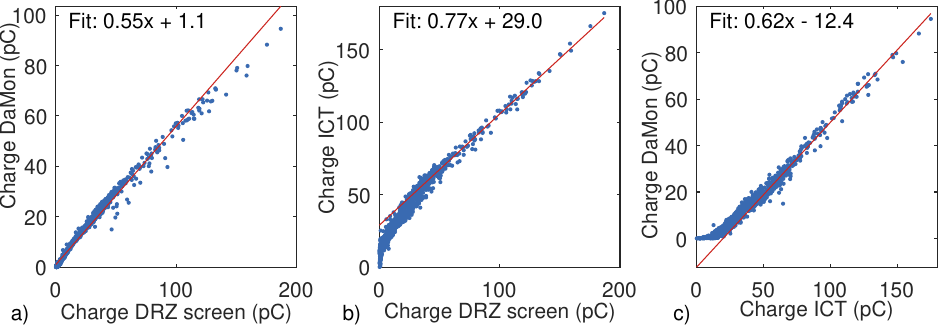}
	\caption[Comparison_largerange]{Comparison of charge measurements and fits. a) DRZ screen vs DaMon, b) DRZ screen vs ICT, c) ICT vs DaMon. For comparisons including the ICT, only datapoints where the ICT measured charges above \SI{50}{pC} where included in the fit.}
	\label{fig:Comparison3diags}
\end{figure*}

By changing the gas density and the laser power in the experiments, it was possible to accelerate electron bunches with charges spanning several orders of magnitude. A comparison of the measured charge for approximately 1500 electron bunches with a variety of charges is depicted in Fig.~\ref{fig:Comparison3diags} together with fits of the measured charge. 

The diagnostics detected charges between \SI{0}{pC} and \SI{200}{pC}. In general a good correlation between the charge measurements was found. However, the individual charge measurement was highly dependent on the diagnostic itself. At low charges, the ICT seems to overestimate the charge and with the exception of only a few data points, charges of around \SI{10}{pC} or more were detected even when both DaMon and DRZ screen measured charges $<$~\SI{1}{pC}, as is visible in Fig.~\ref{fig:Comparison3diags}b and Fig.~\ref{fig:Comparison3diags}c. Towards higher charges ($>$\SI{50}{pC}) a linear correlation between ICT and DRZ screen, as well as ICT and DaMon can be observed. Therefore, only datapoints where the ICT measured charges above \SI{50}{pC} where included for the fits with the ICT.

DaMon and DRZ screen show a linear correlation throughout most of the measurement range, with small differences towards high charges as is visible in Fig.~\ref{fig:Comparison3diags}a. The distribution of the points is also much lower in this panel, indicating both DaMon and DRZ screen are less effected by noise in this setup than the ICT. 
Overall, the following correlations of charge measurements were found using first order polynomial fits:
\begin{equation}
    \begin{aligned}
    	\label{Eq:chargefits}
        Q_{DaMon} & = 0.55 \times Q_{DRZ} + \SI{1.1}{pC} \\
        Q_{ICT} & = 0.77 \times Q_{DRZ} +  \SI{29.0}{pC}\\
        Q_{DaMon} & = 0.62 \times Q_{ICT} - \SI{12.4}{pC}
    \end{aligned}
\end{equation}

The highest charges were detected using the DRZ screen. Similar results were found before and could partly be explained with low energy electrons \cite{Nakamura2016}, which deposit vastly more energy in the screen and therefore lead to a non-linear behaviour of the screen \cite{Glinec2006}. For the presented results, the effect might be further amplified due to the high thickness of DRZ-high and the use of ionisation injection which is likely to produce such low-energy electrons. Another reason for high charge measurements using the DRZ screen could be background radiation such as on-axis bremsstrahlung from the collimator, betatron radiation or laser and/or plasma light as discussed earlier. All of these effects would not be present in the background shots without the laser, and therefore would not be accounted for in the background subtraction. The accuracy of the ICT is brought under question by the non-linear behaviour of the ICT at low charges and previously reported over-estimations of the charge for ICTs which are installed in close proximity to a plasma source \cite{Glinec2006, Hidding2007,Downer2018}. The DaMon measures the lowest charges, detecting nearly half the charge compared to the DRZ screen. In addition to the background effects of the DRZ screen discussed earlier, the differences could also originate from errors in the calibration, e.g. the absolute calibration of the DRZ screen, the cross calibration of the imaging systems using tritium sources or from the calibration of the DaMon. To resolve the differences, an additional comparison of DaMon and a Faraday cup at the presented setup might be required. In previous and recent calibrations at other setups at DESY excellent agreement between DaMon and Faraday cup was achieved \cite{Lipka2011,Lensch2023}.

\section{Summary and Outlook}
In this study, we have presented a comparison of three charge diagnostics, namely the non-invasive diagnostics DaMon and ICT and an absolutely calibrated DRZ screen. 
The DaMon was shown to be capable of detecting electron bunches with low bunch charges of less than \SI{100}{fC}. The imaging system of the DRZ screen detected laser and/or plasma light despite careful shielding, which in experiments with plasma lead to an additional error source. To resolve this, spectral filters should be implemented from the start in absolute calibrations of scintillating screens. The ICT was unable to detect electron bunches with low bunch charge. This could be resolved by using a Turbo-ICT, which has shown to reduce the noise level to approximately \SI{1}{pC} \cite{Nakamura2016}. While this is still not at the level of the DaMon, a comparison of DaMon and Turbo-ICT should be performed in the future to further compare the two diagnostics and their sensitivity to (plasma) EMP.

The influence of strong EMP on DaMon and ICT was further investigated using an active plasma lens. In our tests, an APL in close proximity to the DaMon did not influence its measurement of the electron bunch charge. As such, the DaMon shows great potential for the use as non-invasive charge diagnostic in plasma accelerators.

Furthermore, the performance of the diagnostics over a wide parameter range was investigated. In our measurements, we found that the ICT tends to overestimate the charge of our electron bunches, especially at low charges. These findings are in line with previous results, that indicate an overestimation of the charge by ICTs in plasma accelerators \cite{Glinec2006,Hidding2007}. Throughout the entire measurement range, the charge detected by the DaMon is approximately half of that on the DRZ screen. This could be partly explained by the non-linear response of phosphor screens to low-energy electrons \cite{Glinec2006}, which are expected in ionisation injection and have been identified to play a significant role in charge detection at plasma accelerators in the past \cite{Nakamura2016}. In addition, errors in the absolute screen calibration, the cross calibration using tritium sources or in the calibration of the DaMon could play a role. 
An additional measurement using another absolutely calibrated diagnostic such as a Faraday cup, which previously showed excellent agreement with the DaMon \cite{Lipka2011,Lensch2023} and measurements including cleaning magnets \cite{Nakamura2016} could resolve the differences between DRZ screen and DaMon. The small scattering of datapoints and generally linear correlation between DaMon and DRZ shows that both diagnostics are well suited for charge measurements in plasma accelerators if precisely calibrated. 

Overall, our measurements highlight the difficulties of precise charge measurements in plasma accelerators and compare the cavity-based DaMon to an ICT and a scintillating screen. The Damon provides a non-invasive charge measurement, which is not influenced by EMP and therefore can be implemented in close proximity to the plasma source. Due to its high sensitivity (tens of femtocoulomb) and large dynamic range (seven orders of magnitude) it is well suited for various applications in plasma acceleration. 

\section*{References}


\end{document}